\documentstyle[aps,preprint]{revtex}
\begin{document}
\title{Statistical properties of energy levels of chaotic systems:\\
Wigner or non-Wigner}
\author{Jakub Zakrzewski$^{1,2}$, Karine Dupret$^1$ and
 Dominique Delande$^1$}
\address{$^1$Laboratoire Kastler-Brossel, Tour 12, Etage 1,
Universite Pierre et Marie Curie,\\
4 Place Jussieu, 75005 Paris,
FRANCE}
\address{
 $^2$Instytut Fizyki, Uniwersytet Jagiello\'nski,
 ulica Reymonta 4, 30-059 Krak\'ow, POLAND\footnote{permanent address}.
}
\date{\today}
\maketitle
\begin{abstract}
For systems whose classical dynamics is
chaotic, it is generally believed that the local statistical
properties of the quantum energy levels are well described by
Random Matrix Theory. We present here two counterexamples -
the hydrogen atom in a magnetic field and the quartic oscillator
- which display nearest neighbor statistics strongly different
from the usual Wigner distribution. We interpret the results
with a simple model using a set of regular states coupled to a
set of chaotic states modeled by a random
matrix.
\end{abstract}
\narrowtext

Since the pioneering work of Bohigas et
al \cite{bohigas84}, it has been numerically checked on a wide
variety of systems \cite{bohigas89,haake} that the local 
statistical properties
of a quantum system whose classical dynamics is chaotic, are
well described by Random Matrix Theory. Especially, the
statistical distribution of energy spacing between consecutive
levels - also called Nearest Neighbor Spacing (NNS)
distribution - has been shown to be in excellent agreement
with the spacing distribution between consecutive eigenvalues
of random matrices. For spinless systems with time-reversal
symmetry, the Hamiltonian is real in a suitable basis and the
Gaussian Orthogonal Ensemble (GOE) of random matrices has to be
used. The NNS distribution of this ensemble is very close to
the Wigner surmise \cite{bohigas89,haake}: 
\begin{equation}
P(s)=\frac{\pi s}{2}\exp\left({-\frac{\pi s^2}{4}}\right),
\label{wigner}
\end{equation}
where $P(s){\rm d}s$ measures the probability of
having a spacing in the interval $[s,s+{\rm d}s].$ In order to have
dimensionless quantities, $s$ is here measured in units of
mean level spacing. 

The NNS distribution of specific quantum
chaotic systems generally follows the Wigner prediction for
highly excited states, that is in the semiclassical limit, $\hbar
\rightarrow 0.$ 
 In this paper, we study numerically two simple
systems whose classical dynamics are chaotic, but whose NNS
distributions significantly differ from the Wigner
distribution. These systems have in common that, in some part
of the classical phase space, an adiabatic separation of the
motion along two coordinates is possible. These regions of 
phase space are nevertheless coupled to more irregular
regions and the resulting global motion is classically strongly
chaotic. The surprising result is that, even for very highly
excited states, the approximate adiabaticity is not washed out
in the NNS distribution. Although these examples are clearly
not generic  because of the partial adiabatic
separation, they are of major practical importance. For
example,  a consequence of this study is that the statistical
properties of energy levels of the hydrogen atom in a magnetic
field do not obey the Random Matrix prediction in the vicinity
of the ionization threshold (a region widely experimentally
studied in the last few years \cite{kleppner}),
even for highly excited Rydberg states.

The first
system we consider is the hydrogen atom in a magnetic field
(along the $z $ axis) whose hamiltonian is in atomic units
(neglecting relativistic, spin, QED, finite mass of the nucleus
effects):
\begin{equation}
H=\frac{{\bf p}^2}{2} - \frac{1}{r} + \frac{\gamma^2}{8}(x^2+y^2)
\label{hb}
\end{equation}
where $\gamma$ denotes the magnetic field in
units of $2.35\times 10^5\ {\rm T}.$ 
The paramagnetic term $\gamma L_z/2\ (L_z,$ 
the angular momentum along the $z $ axis is a good
quantum number, together with parity) which can exactly
be taken into account is dropped here. The classical dynamics of
this system depends only on the scaled energy,
$
\epsilon = E \gamma^{-2/3}
$,
and is almost
fully chaotic for $\epsilon > -0.12$.

Various studies of NNS distribution
 \cite{ddprl86,dwpr89} - on spectra obtained
 either at constant magnetic field or at constant scaled
energy - are in good agreement with the Wigner prediction above
$\epsilon =-0.12.$ 
However, careful checks close to the classical ionization
threshold, $E=0$, have not been performed.
 The results presented here have been obtained
in this region for the ($L_z=0,$ odd parity) series of energy
levels. Spectra have been generated at constant magnetic field
by numerical diagonalization of the hamiltonian in a huge
Sturmian basis \cite{ddprl86,dwpr89} 
 using the Lanczos algorithm \cite{ddprl91} which
allows us to compute at low cost the few levels of interest in
a narrow energy band.

The value of $\gamma $ ranges from $9\times 10^{-5}$
 to $6.35\times 10^{-4}$
and the matrix size is up to 77000, from 
which about 50 energy levels
are computed (ranging between the $70^{\rm th}$
 and the $380^{\rm th}$ excited
state, depending on $\gamma ).$
 For each $\gamma $ value, the energy spectrum is
unfolded to determine the average density
 of states and the NNS distribution
further computed. Because we want to study levels in a narrow energy band,
only 5 to 20 spacings were computed for each $\gamma $ value.
478 NNS distributions at various $\gamma $ values
- sufficiently far one from the other to ensure that
they are not correlated - are then collected to
build the distributions plotted in Fig.\ref{1} and Fig.\ref{2}.

For reasons to be explained below,
 the data are collected separately in regions of 
constant $E/\gamma .$ For example, Fig.~\ref{1} 
displays the NNS distribution
for energy levels such that $-0.4 \gamma < E < -0.3 \gamma$ (inset) and the 
corresponding cumulative NNS 
distribution, $\int_0^s{P(x)\ {\rm d}x},$ together with the Wigner
prediction. Clearly, there is a significant deviation; especially,
there are no large spacings.
Fig.~\ref{2} displays the same comparison, but for energy levels
lying closer to the ionization threshold, such that 
$-0.2\gamma < E < -0.1 \gamma .$
The deviation from the Wigner surmise is here even more spectacular,
with the absence of spacings larger than 1.7 mean spacing.

A key point is that the relevant parameter describing the NNS
distribution is the ratio $E/\gamma .$ In Fig.~\ref{2}, we show
two different stretches of $\gamma $ values. Within statistical
errors, the two distributions coincide, although the average
value of $\gamma $ changes by a factor 2.5.
This indicates that the observed deviation persists in the
semiclassical limit $\gamma \rightarrow 0, 
E\rightarrow 0, E/\gamma$ fixed.

The relevant parameter $E/\gamma $ is {\it not} the scaled energy, which
indicates that the deviations observed here are of quantum origin.
A careful study of the phase space structure allows one to understand
the origin of the phenomenon. Indeed, the diamagnetic potential
$\gamma^2\rho^2/8$ confines the motion transverse to the
magnetic field, but not in the magnetic field direction. Hence,
just below the ionization threshold, the electron can explore
the region around the $z$ axis very far from the nucleus. In the
region $|z|\gg \rho ,$ the Coulomb potential $1/\sqrt{\rho^2+z^2}$
is almost equal to $1/|z|,$ leading to the adiabatic-like separation
\cite{kleppner}:
\begin{equation}
H \simeq H_{\rm sep} = H_z+H_{\rho}=\frac{p_z^2}{2}-\frac{1}{|z|}
+\frac{p_{\rho}^2}{2} + \frac{\gamma^2\rho^2}{8}.
\end{equation}
The spectrum of $H_{\rm sep}$ as a sum of the
 1-dimensional Coulomb, $H_z$, and of the harmonic
oscillator, $H_{\rho}$, hamiltonians is
\begin{equation}
E(n_z,n_{\rho}) = (n_{\rho}+\frac{1}{2})\gamma - \frac{1}{2n_z^2}
\label{it}
\end{equation}
with $n_z$ and $n_{\rho}$ integers $(n_z > 0).$

The various $n_z$ Rydberg series associated with different $n_{\rho}$ 
values are coupled by the nonadiabatic term $\frac{1}{|z|}-\frac{1}
{\sqrt{\rho^2+z^2}}.$
 The assumption that the coupling is strong enough to completely
mix the various series allows to model the full hamiltonian by a random
matrix. In the situation discussed here, this assumption is simply
not true. Indeed, the partial density of states in a given 
$n_{\rho}$ channel is $n_z^3.$ As the thresholds for the various
series are different, the corresponding $n_z$ values at a given total 
energy are also different. For example, at $E=-0.5\gamma ,$ the density
of states in the $n_{\rho}=0,1,2...$ series are respectively
proportional to $1,2^{-1.5},3^{-1.5},...$ which means that there are 
much more energy levels of $H_{\rm sep}$ belonging to the $n_{\rho}=0$
series than to the other series.
In other words, the levels belonging to the higher $n_{\rho}=1,2...$
series are not numerous enough to destroy the regularity of the
$n_{\rho}=0$ series, even when the non-adiabatic coupling is strong.

In order to model this situation, we consider a Hilbert space composed
of two subspaces which we baptize ``regular" and ``chaotic".
We construct a model hamiltonian by considering a matrix diagonal in the
regular subspace with equally spaced eigenvalues (this represents
the $n_{\rho}=0$ series discussed above). 
In the chaotic subspace (which represents the strongly mixed
$n_{\rho}=1,2...$ series), we model the hamiltonian by a random matrix.
The coupling of the regular states to the chaotic ones is taken constant
throughout the regular series \cite{coupling}.
Indeed, in the physical system considered here, the Rydberg states
of the $n_{\rho}=0$ series are coupled to the other series close
to the nucleus, where all Rydberg states have almost identical
wavefunctions.

In the limit of large matrices, this model has only two
parameters: the relative weights of chaotic and regular
states (i.e. the ratio of the dimensions of the chaotic and
regular subspaces) and the strength of the coupling. 
More refined models could be used - for example
 taking a nonuniform density of regular states to represent
a Rydberg series - but at the price of adding new parameters and losing
the simplicity of the model. 

We have not been able to deduce analytically the NNS distribution 
for our model and we suspect that it is a very difficult task.
Thus, we have determined it numerically.
It is shown in Figs.~\ref{1} and \ref{2} in comparison with
the NNS distribution for the hydrogen atom in a magnetic field.
The agreement is obviously excellent. Especially, two important
features are well reproduced: first, the absence of large level spacings.
The largest observed spacing precisely
 corresponds to the spacing between two unperturbed states
 of the regular series.
Hence, the largest observed spacing fixes in our model
the ratio of regular to chaotic states. As expected from our
interpretation, this ratio increases from Fig.~\ref{1} 
(ratio=1.35) to Fig.~\ref{2} (ratio=1.72), in agreement
with the increasing density of states in the $n_{\rho}=0$ channel.
This leaves us with a single parameter,
 the strength of the regular-chaotic
coupling, which is adjusted to obtain the best agreement.
The differences between the predictions of the model and the numerical
results on the physical system are of the order of the statistical
fluctuations. This makes us confident that the model
catches the essential part of the physics involved.

 In order to check the validity of the model, we numerically
studied the energy spectrum of a similar system, a quartic oscillator
with hamiltonian:
\begin{equation}
H=\frac{p_x^2+p_y^2}{2}+\frac{x^2y^2}{2}
\label{hx2y2}
\end{equation}

The classical motion is chaotic \cite{bohigas89,pollak,x2y2,tomsovic}. 
Although the
particle can escape classically along $x$ or $y$ axis whatever
its positive energy, it has been proven \cite{bohigas89} that the quantum
energy spectrum is purely discrete.

The hamiltonian~(\ref{hx2y2}) has a $C_{4v}$ symmetry, leading
to four non-degenerate series of energy levels, which we 
label EEE, EEO, OOE and OOO according to Ref.~\cite{x2y2} 
(E means even, O means odd, the first two letters refer
to the $x\rightarrow -x$ and $y\rightarrow -y$ symmetries, 
the third letter to the $x\leftrightarrow y$ symmetry), plus
a twofold degenerate series (EO and OE) that we will not consider here.

The energy levels are computed through diagonalization
of the hamiltonian in a harmonic oscillator basis. Matrices 
of size up to 922000 are used to produce the 4 independent
sets of energy levels. The spectra are unfolded \cite{tomsovic,ddjz}
and the NNS distributions computed. We find that the EEE and EEO
series have similar statistical properties, as well as the
OOE and OOO series. We thus collect the corresponding data in two sets.
The cumulative NNS distributions are shown in Figs.~\ref{3} and \ref{4},
together with the best fits using our regular-chaotic model.
Again, the agreement is very good in both cases. The phenomena already
observed for the hydrogen atom in magnetic field,
namely the lack of large spacings and strong deviations from the Wigner
distribution, are also observed here.

As soon as the particle escapes either in the $x$ or the $y$ direction,
an adiabatic separation in coordinates $x$ and $y$ is possible. For example,
in the $x$ direction, the particle oscillates rapidly around the
$y=0$ equilibrium position while moving slowly along $x.$ The energy
spectrum is thus composed of interacting series, each series
corresponding to a well defined $n_y$ quantum number
of the harmonic oscillator along $y$ and various states along the
$x$ direction. For even-$y$ parity, the series $n_y=0$ has a much
higher density of states than the other series $n_y=2,4,....,$
leading to strong deviations from the Wigner NNS distribution
(see Fig.~\ref{3}). For odd $y$-parity, 
the coupled series are $n_y=1,3,5...$
The $n_y=1$ series is dominant over the other ones, but the effect is less
pronounced (see Fig.~\ref{4}): indeed, the $n_y=1$ wavefunction vanishes
on the $y=0$ axis and cannot ``explore" the motion along $x$ as much as the 
$n_y=0$ state.
We have also checked that the numerically obtained distributions 
do not depend on the energy, which means that it should
persist for highly excited states in the semiclassical limit.

Finally, we want to stress that the observed deviations have
an intrinsic {\it quantum} origin. It is because few series
interact - one being dominant - that Wigner distribution is not observed.
For the hydrogen atom in a magnetic field, $E/\gamma $ has to be
kept close to 0. This means that, at constant scaled energy
$\epsilon =E\gamma^{-2/3}$ in the semiclassical limit
$E\rightarrow 0,\gamma \rightarrow 0,$ the ratio $E/\gamma $ tends
to $-\infty$ and the Wigner distribution is recovered.
When speaking about ``semiclassical limit", one has to
be careful, indicating precisely how the various
quantities tend to 0.

Although the situation discussed in this letter seems to be
exceptional, it has a tremendous practical importance, especially
in atomic physics. Indeed, the Coulomb interaction has an
infinite range and
may easily produce such situations where an adiabatic
separation is possible in some part of phase space.
Let us give two simple examples. For the hydrogen atom
in a magnetic field close to the ionization
threshold, either below or above, one is always close
to a Landau threshold $(n_{\rho}+1/2)\gamma$ where a Rydberg
series converge to. Hence, strong deviations from random
matrix predictions are expected everywhere in this region,
either for bound states or resonances \cite{karine}.
This has in fact been observed experimentally 
\cite{kleppner}. These authors discovered windows where one
Rydberg series appear to be accidentally decoupled from the other ones,
giving rise to a locally regular series. Although this decoupling
is exceptional throughout the spectra, our statistical observations rely
on the same phenomenon. 

The second example is even more spectacular, and probably quite obvious for 
most atomic physicists. The helium atom is a mainly chaotic three-body 
system. Nevertheless, below every single
 ionization threshold, one observes regular
Rydberg series whose NNS distribution is far from Wigner (it is actually
a $\delta$-peak at $s=1$). This is just the extreme limit of our model
when one series has a much larger density of states than all the other
ones (associated with perturbers in 
the Rydberg series) and the chaotic subspace almost
disappears.

J.Z. acknowledges partial support of KBN, grant P302 102 06.
Laboratoire Kastler-Brossel, de l'Ecole
Normale Sup\'erieure et de l'Universit\'e Pierre et Marie Curie, is 
Unit\'e Associ\'ee 18 du Centre National de la Recherche Scientifique.

\begin{figure}
\caption{Cumulative nearest neighbor spacing (NNS) distribution
for the hydrogen atom in a magnetic field close to the ionization 
threshold. The NNS distribution itself in shown in the inset.
The data are collected for 3271 spacings in the energy band
$[-0.4\gamma ; -0.3\gamma ]$ for 478 values of $\gamma .$ It deviates
noticeably from the Wigner prediction, Eq.~(\protect{\ref{wigner}}) 
shown as a dotted line. The dashed line represents the prediction
of a simple model where a series of equally spaced levels
is coupled to a GOE matrix (ratio of regular levels to chaotic
levels = 1.35).}
\label{1}
\end{figure}

\begin{figure}
\caption{Same as Fig.~1, but for states in the energy band $[-0.2\gamma ;
-0.1\gamma ].$ The sample of 
4293 spacings is divided in two independent samples (differing by a
factor 2.5 in the average value of $\gamma ).$ The two corresponding
solid lines almost coincide, which proves that the deviation from
the Wigner distribution persists in the $\gamma \rightarrow 0$ limit
where higher and higher excited states are considered. The ratio
of regular levels to chaotic levels in the model is 1.72.}
\label{2} 
\end{figure}

\begin{figure}
\caption{Cumulative nearest neighbor spacing distribution 
for the quartic oscillator, Eq.~(\protect{\ref{hx2y2}}). 
Only the two series
of states symmetric with respect to the $x$ and $y$ axis are shown here.
There are 381 level spacings included (after removing the first 100 levels
from each subset; the number of removed  
states does not change the shape of the distribution).
The dashed line represents the prediction of our regular-chaotic model
(ratio of regular levels to chaotic levels = 2.64). The inset shows
the NNS distribution itself.}
\label{3}
\end{figure}

\begin{figure}
\caption{Same as Fig.~3, but for 
the two series of states antisymmetric
with respect to the $x$ and $y$ axes. There are 924 level spacings 
included. The dashed line represents the prediction of our
regular-chaotic model (ratio of regular levels to chaotic levels = 0.82).}
\label{4}
\end{figure}


\begin{references}
\bibitem{bohigas84}
O. Bohigas, M.J. Giannoni and C. Schmit, Phys. Rev. Lett. 52, 1 (1984). 

\bibitem{bohigas89} O. Bohigas in 
{\it Chaos and quantum physics},
edited by M.-J. Giannoni, A. Voros and J. Zinn-Justin, 
Les Houches Summer School, Session LII (North-Holland, Amsterdam, 1991).

\bibitem{haake}  F.~Haake, {\it Quantum Signatures of Chaos}
                    (Springer, Berlin 1991).

\bibitem{kleppner} C. Iu, G.R. Welch, M.M. Kash, K. Hsu,
and D. Kleppner, Phys. Rev. Lett. {\bf 63}, 1133 (1989).

\bibitem{ddprl86}
D. Delande and J.C. Gay, Phys. Rev. Lett., 57, 2006 (1986);
 D. Wintgen and H. Friedrich, Phys. Rev. Lett.,
57, 571 (1986); G. Wunner, U. Woelk, I. Zech, G. Zeller,
 T. Ertl, P. Geyer, W. Schweizer and P. Ruder,
Phys. Rev. Lett., 57, 3261 (1986);


\bibitem{dwpr89} H. Friedrich and D. Wintgen, 
Phys. Rep. {\bf 183}, 37 (1989).

\bibitem{ddprl91} D. Delande, A. Bommier, J.C. Gay, 
Phys. Rev. Lett. {\bf 66}, 141
(1991).

\bibitem{coupling}
As the hamiltonian in the chaotic subspace is modelled by a GOE
matrix, all states
of the chaotic subspace play equivalent role (the ensemble 
is invariant under any orthogonal transformation).
Thus the repartition of the regular-chaotic coupling
over the chaotic states does not matter. 


\bibitem{pollak} B.~Eckhardt, G.~Hose and E.~Pollak,
Phys.\ Rev.\ A{\bf 39} (1989) 3776.

\bibitem{x2y2} C.~C.~Martens, R.~L.~Waterland, and W.~P.~Reinhardt,
J.\ Chem.\ Phys.\ {\bf 90} (1989) 2328.

\bibitem{tomsovic} S.~Tomsovic, J. Phys. A {\bf 24}, L773 (1991).

\bibitem{ddjz} D.~Delande and J.~Zakrzewski, to be published.

\bibitem{karine} K.~Dupret,  D.~Delande and J.~Zakrzewski,
 to be published.

\end{references}
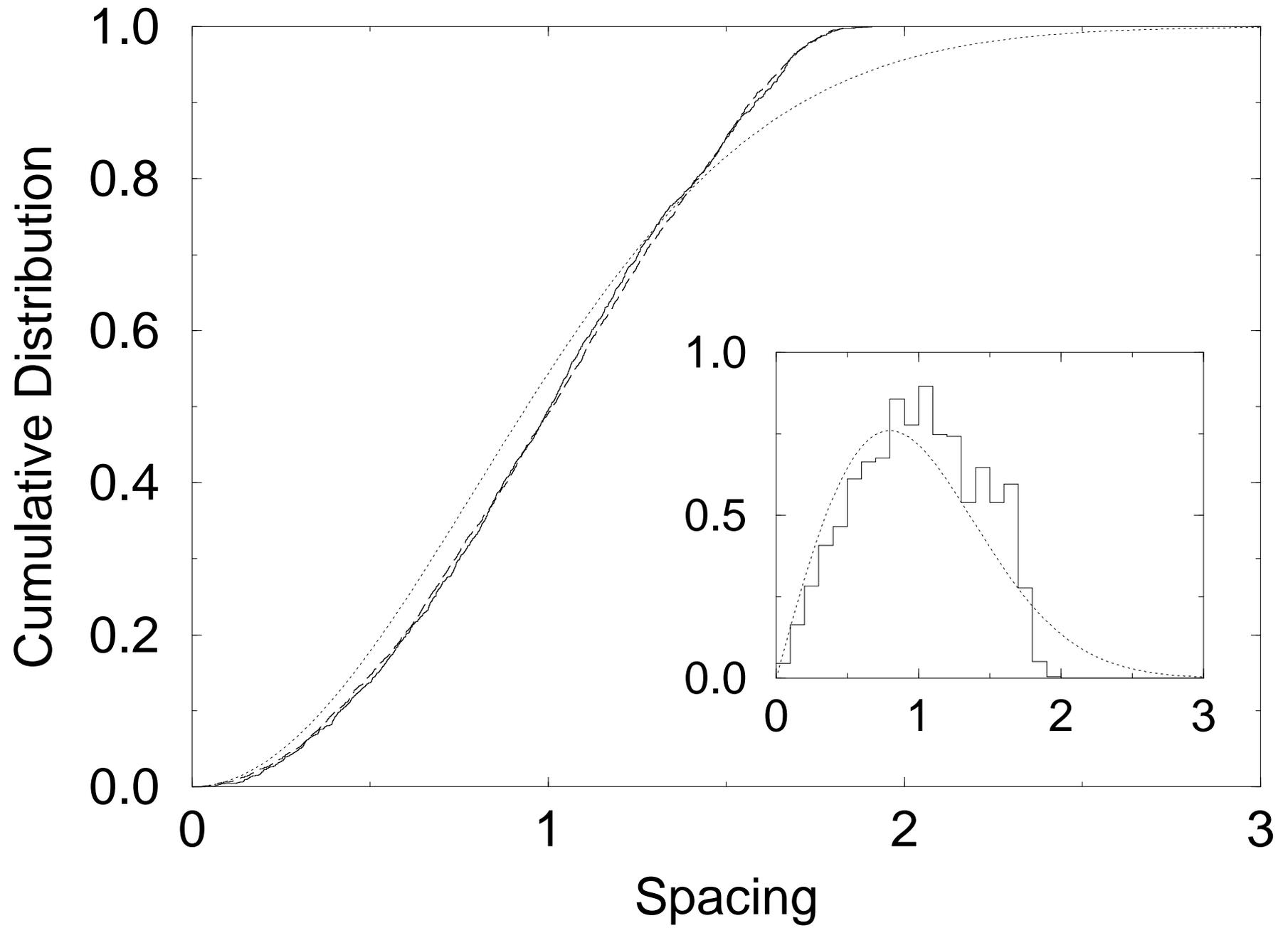
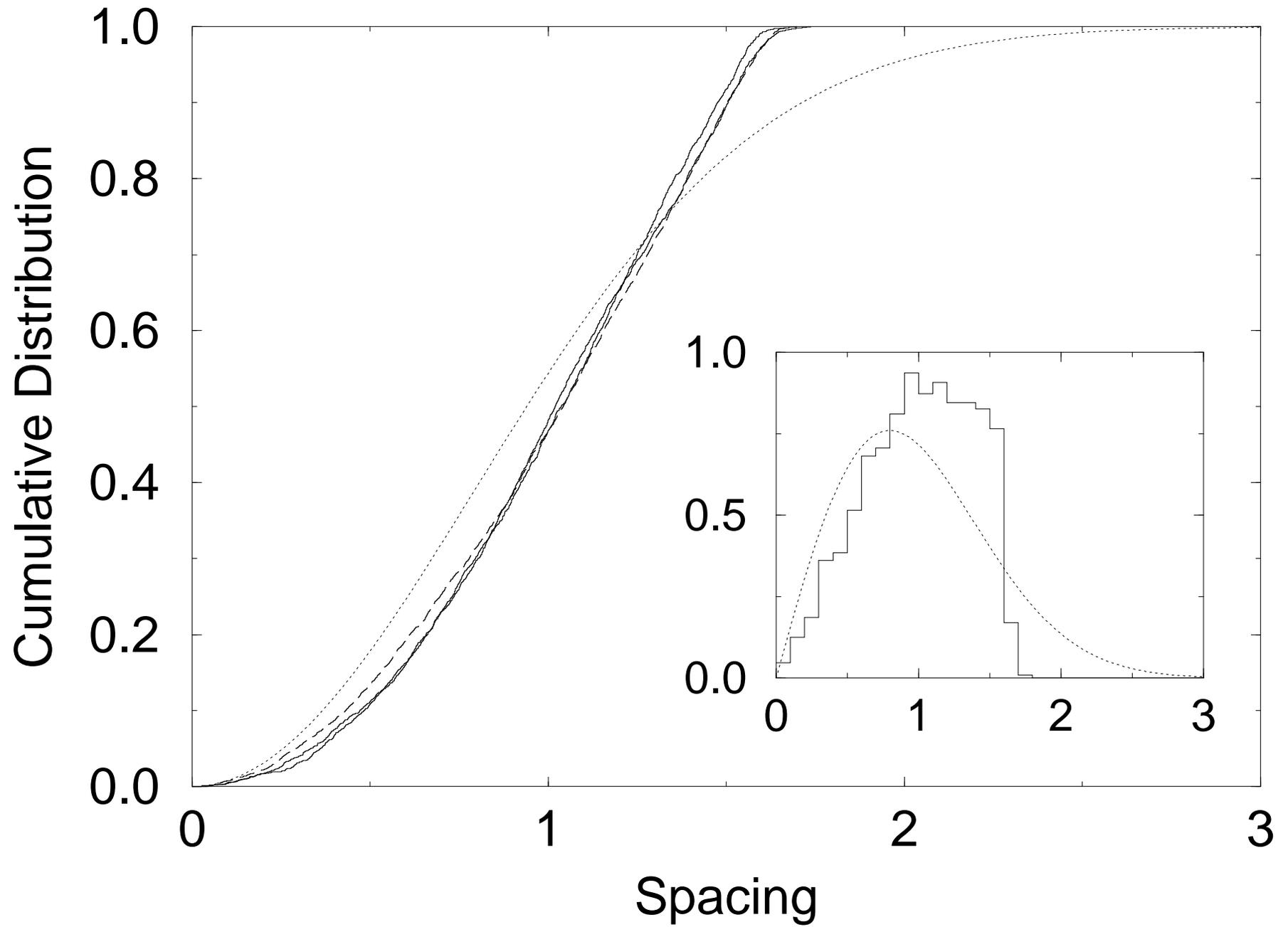
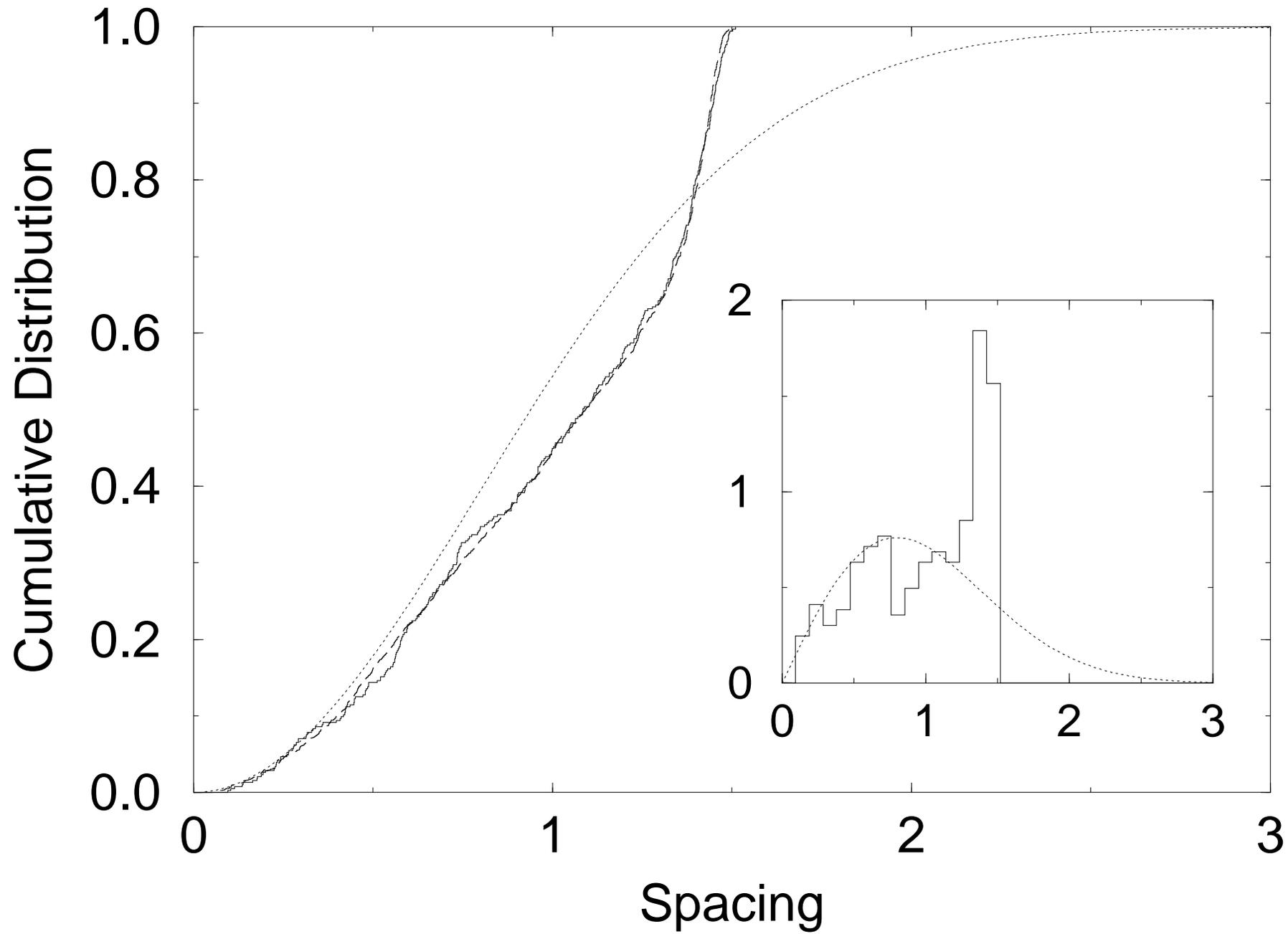
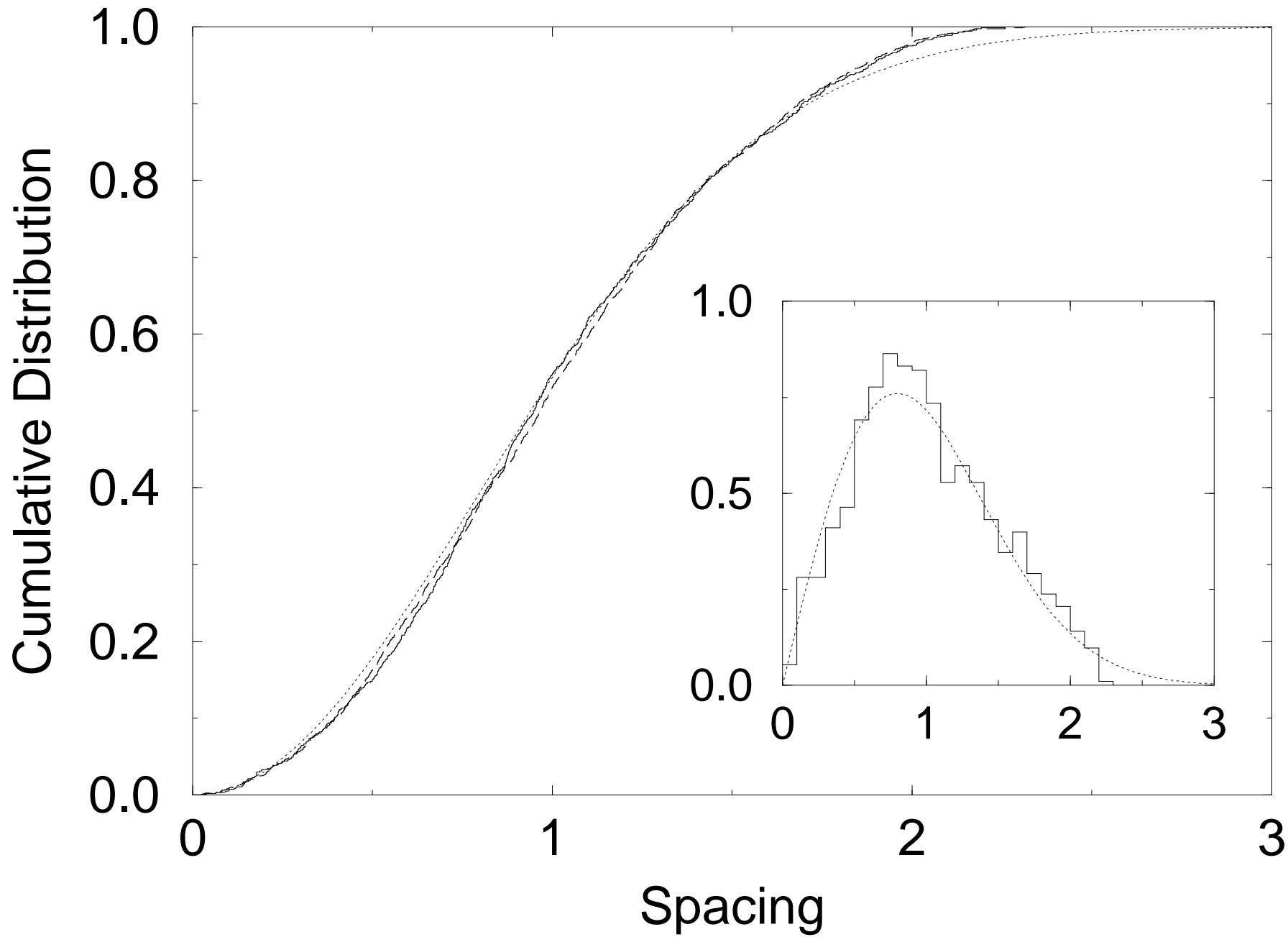
\end{document}